\documentclass[manuscript]{acmart}
\usepackage[hyphenbreaks]{breakurl}
\AtBeginDocument{%
  \providecommand\BibTeX{{%
    \normalfont B\kern-0.5em{\scshape i\kern-0.25em b}\kern-0.8em\TeX}}}

\copyrightyear{2022}
\acmYear{2022}
\setcopyright{rightsretained}
\acmConference[FAccT '22]{2022 ACM Conference on Fairness, Accountability, and Transparency}{June 21--24, 2022}{Seoul, Republic of Korea}
\acmBooktitle{2022 ACM Conference on Fairness, Accountability, and Transparency (FAccT '22), June 21--24, 2022, Seoul, Republic of Korea}\acmDOI{10.1145/3531146.3533163}
\acmISBN{978-1-4503-9352-2/22/06}

\begin{document}

\title{Regulating Facial Processing Technologies: Tensions Between Legal and Technical Considerations in the Application of Illinois BIPA}

\author{Rui-Jie Yew}
\email{rjy@mit.edu}
\authornote{Work done in part while author was an intern at Sony AI.}
\affiliation{%
  \institution{Massachusetts Institute of Technology}
  \city{Cambridge}
  \state{MA}
  \country{USA}}

\author{Alice Xiang}
\email{alice.xiang@sony.com}
\affiliation{%
  \institution{Sony AI}
  \country{USA}}

\renewcommand{\shortauthors}{Yew and Xiang}
\renewcommand{\shorttitle}{Regulating Facial Processing Technologies: Tensions in Applications of Illinois BIPA}
\begin{abstract}
Harms resulting from the development and deployment of facial processing technologies (FPT) have been met with increasing controversy. Several states and cities in the U.S. have banned the use of facial recognition by law enforcement and governments, but FPT are still being developed and used in a wide variety of contexts where they primarily are regulated by state biometric information privacy laws. Among these laws, the 2008 Illinois Biometric Information Privacy Act (BIPA) has generated a significant amount of litigation. Yet, with most BIPA lawsuits reaching settlements before there have been meaningful clarifications of relevant technical intricacies and legal definitions, there remains a great degree of uncertainty as to how exactly this law applies to FPT. What we have found through applications of BIPA in FPT litigation so far, however, points to potential disconnects between technical and legal communities. This paper analyzes what we know based on BIPA court proceedings and highlights these points of tension: areas where the technical operationalization of BIPA may create unintended and undesirable incentives for FPT development, as well as areas where BIPA litigation can bring to light the limitations of solely technical methods in achieving legal privacy values. These factors are relevant for (i) reasoning about biometric information privacy laws as a governing mechanism for FPT, (ii) assessing the potential harms of FPT, and (iii) providing incentives for the mitigation of these harms. By illuminating these considerations, we hope to empower courts and lawmakers to take a more nuanced approach to regulating FPT and developers to better understand  privacy values in the current U.S. legal landscape.
\end{abstract}




\maketitle

\section{Introduction}
In 2021, Facebook’s \$650 million Illinois Biometric Information Privacy Act (BIPA) settlement regarding its use of facial recognition technology~\citep{fb22621} and its subsequent deletion of more than a billion of its users’ face templates~\citep{pesenti2021} has reignited focus on legal privacy considerations in the building and deploying of facial processing technologies (FPT).  FPT encompasses (i) facial recognition technology, which is used to identify or verify identities based on facial images, (ii) face detection technology, which detects the presence of and key points on the face, and (iii) facial analysis technology, which can be used to glean demographic information and emotions~\citep{raji2020about}. FPT in commercial establishments~\citep{gershgorn2021}, exam proctoring~\citep{clark2021}, and even apartment complexes~\citep{durkin2019}, has enabled unwelcome surveillance and has disproportionately harmed Black and Brown communities in the United States, leading to wrongful arrests~\citep{ryan-mosley2021}~\citep{stokes2020} and the targeting and tracking of immigrants~\citep{chappell2019}. The seminal work of Buolamwini and Gebru~\citep{buolamwini2018} brought increased attention to the harms that result from disparities in the accuracy of industry gender classification algorithms, with darker-skinned, feminine-presenting individuals having the lowest accuracy rate across demographic groups. As a result, cities such as Somerville, Massachusetts~\citep{somerville2019} and the state of Maine~\citep{maine2021} have prohibited the use of facial recognition technology in most contexts by law enforcement and by the government. New York City also introduced its biometric privacy law in July of 2021, requiring notice of biometric technology use, including FPT use, in commercial establishments such as grocery stores and restaurants~\citep{nystateassembly2021}. While these state and city policies have banned the use of FPT in certain scenarios, there continues to be the development and use of FPT in a wide range of contexts that are not implicated by these state and city FPT policies~\citep{simonite2021}. Instead, they are regulated by broader state information privacy laws, like the California Consumer Privacy Act (CCPA) and Illinois BIPA. Even though there are several U.S. state privacy laws that regulate biometric privacy~\citep{txlegislature2017,calegislature2018}, litigation around FPT in the United States is currently heavily centered on BIPA. 
Enacted in 2008, Illinois BIPA is one of the oldest laws in the United States that governs biometric privacy~\citep{illegislature2008}. Because of BIPA’s longevity and its unique position as the only biometric privacy law in the United States allowing private right of action for the nonconsensual processing of biometric information~\citep{hartzog2021}, a significant number of lawsuits that allege violations of the act have progressed, and BIPA is responsible for the largest settlement amounts by tech companies for their development and deployment of FPT~\citep{fb22621, shutterfly090921, tiktok093021}.  BIPA governs the collection, possession, use, distribution, and sale of biometric identifiers and biometric information. Under the policy, a ``biometric identifier'' includes a ``scan of face geometry'' and ``biometric information'' includes any information that is based on an individual’s biometric identifier~\citep{illegislature2008}. 

For FPT, BIPA litigation has centered around two forms of data that have been considered scans of face geometry by courts—face templates and facial landmarks. Face templates, or faceprints, are unique representations of people’s faces which can be used to measure differences between people’s eyes, noses, and lips, and can also encode features like skin tone and texture. A person’s face template can also be learned and extracted from a variety of head poses, making them representative of a person’s facial identity, rather than just informative about certain parts of a person’s face. Face templates were used as part of Facebook’s Tag Suggestions, which has since been replaced by its facial recognition feature~\citep{pesenti2021}. Facial landmarks, on the other hand, can be thought of as points that mark the positions of features of your face, such as your eyes, nose, and lips. The level of information that facial landmarks provide can differ dramatically based on how many markers constitute them. Facial landmarks can constitute just five landmark markers that are informative of a particular portion of a subject’s face~\citep{king2017} or over 90 markers~\citep{snapchatlensstudio} that map out the subject’s entire face. Facial landmarks may be extracted as part of the face alignment step for facial recognition~\citep{kumar2020} and in face tracking applications that track positions of facial features in real-time~\citep{appledeveloper}. Despite facial landmarks not necessarily capturing a subject’s facial identity, they can still be used to infer sensitive data, such as emotional information ~\citep{tautkute2018}. 

Because BIPA governs facial biometric data, not specific forms of FPT, and because of BIPA's private right of action, the policy has generated cases against a tremendous breadth of different types of FPT within different contexts. Even beyond the litigation that BIPA informs, the policy has created incentives for companies to preemptively disable certain features that require facial data in Illinois. For example, Google disabled its Arts \& Culture selfie feature in Illinois in 2018~\citep{langone2018}, Google Nest does not enable the familiar face detection feature in Illinois~\citep{googlefacedetection}, and Sony does not sell its Aibo dog robot to customers in Illinois~\citep{sony2018}.

Our contributions in this paper are two-fold. First, we provide an overview of how legal standing has been determined in FPT BIPA cases. Legal standing is important to develop an understanding of the harms associated with FPT that are recognized as legally cognizable. Second, we lay out four main FPT considerations where there are disconnects between technical and legal communities in the application of BIPA. It is important to note that, under BIPA, there remains a lack of specific regulatory guidance for FPT and a significant degree of legal uncertainty. This is because (a) most companies settle cases before decisions about technical intricacies and clarifications of legal definitions can be made and (b) Illinois BIPA was enacted in 2008, before the widespread development and use of computer vision technologies. The picture we paint is one where there are important nuances in the technical operationalization of the policy for FPT that are not be captured in BIPA or the courts’ applications of it, as well limitations of applying solely technical methods to achieve legal privacy values. We hope that this paper’s mapping of insights from different U.S. FPT cases and analysis of relevant legal and technical distinctions will empower courts and lawmakers to take a more nuanced approach to regulating FPT, as well as enable developers to better understand the current legal landscape.

\section{Legal Standing and Harms}
Legal standing in BIPA cases provides a window into what constitutes a legally cognizable privacy harm in the context of FPT. Plaintiffs in privacy lawsuits have generally focused on satisfying the federal standing requirement that the violation of the policy has led to an ``injury in fact'', or that the violation of the policy has brought the plaintiffs a ``concrete, particularized harm'', where particularized is taken to mean a harm that is specific to an individual~\citep{lujan1992}. Hartzog~\citep{hartzog2021} notes that, unlike for other U.S. privacy laws, several courts have recognized violations of BIPA as a legally cognizable injury in itself (sufficiently concrete and particularized to grant legal standing). While this has been the case for some alleged violations of BIPA as a result of FPT~\citep{fb22621}, in practice for FPT litigation, courts have taken a variety of stances regarding which harms qualify as sufficiently particularized and concrete. In this section, we introduce and analyze how privacy harms and standing have been presented in FPT court opinions. In Section~\ref{section:particularized}, we provide an overview of determinations of particularized harms from violations of BIPA, with a focus on BIPA's profit clause. We discuss how courts have determined whether harms arising from BIPA's profit clause are sufficiently particularized, and we argue that this determination does not give enough weight to the harms that might come from the procurement of biometric data and the fundamentality of that data in the development of FPT. In Section~\ref{section:concrete}, we discuss the harms that have been recognized as concrete and note a lack of court consideration of discrimination harms.

\subsection{Particularized Harms}\label{section:particularized}
In this section, we discuss how courts have assessed whether FPT harms arising from alleged violations of various components of BIPA are sufficiently particularized to confer standing, focusing on BIPA's profit clause. For determinations of particularized harms and standing for violations of BIPA's disclosure and consent clauses, we consider \textit{Hazlitt v. Apple}~\citep{apple111220} and \textit{Vance v. IBM}~\citep{ibm091520}, cases involving biometric identifiers in face grouping technology and training data, respectively. For both of these cases, harms arising from violations of the disclosure policy (considered a duty to the public) were not sufficiently alleged for the violation to be upheld, while harms arising from alleged violations of the consent policy (considered a duty owed to an individual) were sufficiently alleged~\citep{apple111220,ibm091520}. The District Court in Apple's case took care to note that violations of duties to the public can still be considered particularized, if sufficiently alleged. Overall, these court opinions suggest differences in privacy duties for different portions of BIPA in the context of FPT, as well as different standards for how violations of these duties should be alleged in order to be upheld.

BIPA’s profit clause has generated the most interesting implications about particularized harms regarding FPT, with courts having the opinion that, since companies do not benefit from a specific individual’s biometric data, but from the company’s FPT as a whole, violations of the profit provision do not confer standing~\citep{apple111220,microsoft031521}. In the determination of whether a harm is sufficiently particularized as a result of alleged violations of the profit policy, there has been a sole focus on biometric data dissemination in the final product, rather than the data's use in the technology's development~\citep{microsoft031521, amazon041421, motorola010821}. For example, in \textit{Vance v. Microsoft}, the plaintiffs’ argument that their biometric data improved Microsoft’s facial recognition technologies did not confer them standing because the allegations did not ``establish that Microsoft disseminated or shared access to biometric data through its products''~\citep{microsoft031521}. However, alleged violations of the profit clause have been upheld when biometric data was used as part of an underlying searchable face database or for face matching, such as in \textit{Flores v. Motorola Solutions} and \textit{Vance v. Amazon} ~\citep{motorola010821, amazon041421}. In \textit{Flores v. Motorola Solutions}, the plaintiffs alleged that their biometric data contributed to Motorola Solutions’ searchable face database, while, in \textit{Vance v. Amazon}, the plaintiffs emphasized the potential for their biometric training data to improve Amazon Rekognition, Amazon’s machine learning API that enables face matching. The District Court for the Western District Court of Washington in Amazon’s case noted that, similarly to Motorola Solution’s case, ``the biometric data [in Amazon’s FPT] is itself so incorporated into Amazon's product that by marketing the product, it is commercially disseminating the biometric data''~\citep{amazon041421}. But \textit{how} the biometric face data was used in the two cases likely differed. In \textit{Flores v. Motorola Solutions}, the plaintiff’s biometric data was extracted as part of a mugshot database and would likely be checked against through every use of the product. In \textit{Vance v. Amazon}, it is less clear if the plaintiff’s biometric data was singled out and disseminated in the same way. Since the plaintiffs’ biometric data was extracted from IBM’s Diversity in Faces dataset, it is likely that the plaintiff’s biometric data in Amazon's case was used to train Amazon’s FPT, including Amazon Rekognition, but not included as part of an underlying database that Amazon provides to its consumers.  This then makes how the biometric data was used in \textit{Vance v. Amazon} more similar to how biometric data was used in \textit{Vance v. Microsoft}, where the profit allegations were dismissed.

At the same time, the capabilities of most corporate FPT result from the biometric training data of hundreds of thousands, if not millions, of individuals, with it being difficult to determine how exactly an individual’s biometric data influences an FPT system more broadly. Even if the biometric data of a specific individual is not  disseminated through every use of FPT, the data is still crucial to fundamental FPT functionality. In fact, biometric face training data is so fundamental to FPT that attacks to certain corporate machine learning APIs have been found to reconstruct an individual’s face from the training data given their name~\citep{fredrikson2015}. Thus, in some sense, even for some forms of FPT that are not searchable face databases or intended to allow for face-matching, a skilled user can single out an individual’s biometric data in a way that might implicate BIPA profit protections under current court opinions, after all. The ease with which facial biometric information can be obtained and disseminated as well as the fundamental function it contributes to FPT differentiates it from other forms of biometric information, such as fingerprint data. This ease of collection has led to the covert collection of face biometric data, such as through surveillance footage~\citep{fussell2019}. Biometric face data's contribution to FPT's fundamental functionality has also created growing demand for that data: there are companies that provide synthetic training data~\citep{wood2021} and tasks for Amazon Mechanical Turkers to upload images of themselves~\citep{sannon2019}. Consequently, requiring the singling out of biometric data in the use of FPT to satisfy a sufficiently particularized violation of the profit clause does not capture the harm a company can cause through the procurement of an individual’s FPT biometric data. It matters, too, how the biometric data was obtained and leveraged in the company’s development of FPT and the importance of the data’s contribution to the FPT’s overall functionality.

In Microsoft's case, where the profit allegation was dimissed, the court noted that ``BIPA was not intended to stop all use of biometric technology; instead, it set a standard for the safe collection, use, and storage of biometrics, including protecting against the public's main fear that their biometric data would be widely disseminated. Section 15(c) [the profit clause] achieves that goal by prohibiting a market in the transfer of biometric data, whether through a direct exchange—sale, lease or trade—or some other transaction where the product is comprised of biometric
data.'' But is not Microsoft's, as well as Amazon's, procurement of the plaintiff's biometric data from IBM's Diversity in Faces dataset telling of an existing market that disseminates biometric training data? Considering biometric data's contribution to profit in the development of FPT rather than just its dissemination in the FPT's deployment would increase corporate responsibility around how training data is sourced and provide meaningful protections against the privacy intrusions that the subjects of that data might have experienced through their data's covert collection.

\subsection{Concrete Harms}\label{section:concrete}
BIPA seeks to protect biometric data in financial transactions and address public hesitancy surrounding biometric information use. The policy intends to prevent privacy harms that arise from the compromising of biometric data, such as identity theft and the theft of financial and other personal information, which have been increasingly tied to biometrics. These privacy harms can largely be characterized as economic and autonomy privacy harms under Citron and Solove’s typology of harms~\citep{citron2022}. Yet, they do not entirely account for the intrusions that come particularly with FPT, which have, under the same typology, also imparted significant discrimination harms~\citep{ryan-mosley2021, stokes2020}. We look to \textit{Rivera v. Google} as a case where the court dismissed all alleged violations of BIPA because they did not pose concrete economic and autonomy harms. With \textit{Patel v. Facebook}’s BIPA settlement and the recognition of an injury-in-fact, we note that litigation is moving towards a general recognition that BIPA violations confer autonomy harms, but there remains a lack of discrimination privacy harm considerations that come specifically with FPT.

In \textit{Rivera v. Google}, Google was sued for its non-consensual collection of face templates for face grouping in Google Photos. The District Court for the Northern District of Illinois, Eastern Division dismissed the motion for lack of standing on all counts, claiming that ``there is no legislative finding that explains why the absence of consent gives rise to an injury that is independent of the risk of identity theft''~\citep{google122918}. With this opinion, there is the implicit assumption that concrete privacy harms under BIPA only arise from potential external access to biometric data and security risk, not necessarily from the non-consensual collection of data from private corporations that stays internal. The court’s focus on the harm of data breach aligns with BIPA’s legislative intent to protect against identity theft, as well as the policy’s focus on how biometric information is collected and disclosed rather than used, echoing the argument presented in~\citep{kugler2019} that courts have determined harms resulting from a violation of BIPA by asking what harms are introduced as a result of not disclosing that biometric information was collected, rather than by asking how people were or could be harmed by how the information was used. 

In contrast, in \textit{Patel v. Facebook}, the Court of Appeals for the Ninth Circuit ruled that Facebook’s violation of BIPA constituted an ``injury in fact'', already  concrete and particularized. The Ninth Circuit’s analysis of this case noted the potential for face templates collected by Facebook to be used in more surveillant contexts, both internally by Facebook and by others externally to, for example, identify a user in a ``surveillance photo taken on the streets or in an office building''~\citep{facebook080819}. This led the court to uphold the motions on all counts, acknowledging that ``a violation of the Illinois statute injures an individual’s concrete right to privacy'', that even the internal collection of biometric data qualifies a concrete injury~\citep{facebook080819}.

Certainly, courts’ broad recognition of harms arising from BIPA as concrete can place meaningful guardrails on the development of FPT that target individuals, but this general recognition may not address the severity of the privacy harms that disproportionately affect marginalized groups. In the complaint for \textit{Flores v. Motorola Solutions}, the plaintiffs noted that, to be included in the mugshot database from which Motorola Solutions extracted biometric identifiers, one merely had to have been arrested, even if the arrest was made in error. With Motorola Solutions’ face database and technology being sold to and used especially by law enforcement, marginalized populations are more easily found through the ``perpetual line-up'' that such FPT enables~\citep{garvie2016}. The court acknowledged that individuals in publicly published photographs also deserve privacy protections for their biometric data but did not acknowledge the additional harm of increased findability that could come with those photographs being mugshots. Similarly, in \textit{Vance v. Amazon}, where the feature of face matching with images from a ``diverse'' dataset was emphasized in the court opinion, there was no discussion of discrimination harms that could come with the targeting of underrepresented individuals in the procurement of such a dataset~\citep{motorola010821}. The focus of court opinions on economic and autonomy harms is unsurprising given BIPA’s legislative intent and the breadth of technology it governs, but it has meant correspondingly little consideration for the discrimination harms that FPT development and deployment has the potential to impart.

\section{Considerations}

BIPA litigation has highlighted a number of tensions in privacy values between technical and legal communities. We center technical communities heavily involved in the development of FPT, focusing on private industry, and we center legal communities involved in the application of the policy, focusing on courts. By tensions, we mean areas where the technical operationalization of BIPA may create unintended and undesirable incentives for FPT development, as well as areas where BIPA litigation can serve to illuminate the limitations of solely technical methods in achieving legal privacy values. In this section, we discuss four relevant points. In Section~\ref{section:controlandbatteryconsumption}, we note that, while the technical community may consider on-device data processing as an avenue to achieve privacy preservation, the legal community emphasizes control over data processing even if that processing occurs on-device. In Section~\ref{section:consent}, we note that, while BIPA places a strong focus on written consent in the development of privacy-preserving FPT, there remains ambiguity about how consent should be prioritized within the technical community. In Section~\ref{section:identifiability}, we note that, while the technical community has made efforts to increase privacy protections in facial images, BIPA explicitly excludes its protections for facial images, which, taken to the extreme, could push against those very efforts. Lastly, in Section~\ref{section:purposeofuse}, we note that, while FPT purpose of use has become increasingly consequential in design and deployment decisions, the way that technology resulting from biometric identifiers or information is used is not considered as part of BIPA or court opinions.

\begin{table}[ht]\label{table:tensions}
\caption{\label{table:tensions}: Tensions between technical and legal communities.}
\begin{tabular}{ p{3cm} p{6.5cm} p{6.5cm}  }
\toprule
\textbf{Theme}&\textbf{Values in FPT Development}&\textbf{Values in BIPA Litigation and Policy}\\\hline\\
Control and Battery Consumption & On-device data processing is a privacy-preserving mechanism that can absolve data of the privacy protections that a company might otherwise be responsible for if the data was processed on its servers.& User control over data is crucial even when processing occurs locally; when users do not have control over data processed on-device, the company may possess that data under BIPA. The battery power consumed when data is processed on devices also is considered a legally cognizable harm under BIPA.
\\\\
Consent	& There is division in the prioritization of consent among other values, such as fairness and the reliance on other legal mechanisms, such as licenses in the acquisition of biometric information and facial images. & BIPA and BIPA litigation place a strong focus on consent, both when companies already have a relationship with the subjects of biometric information (for example, when the subjects are on the company's image-sharing platform) and when companies do not have a relationship with the subjects of biometric information and acquire it another way.
\\\\
Identifiability	& Face obfuscation and blurring in the de-identification of facial images are important in preserving privacy. Additionally, the degree to which the subject of a particular technology resulting from biometric data or of the biometric data can be used to identify that subject is an important consideration.	& BIPA explicitly excludes photographs from its definition of biometric identifiers and biometric information, hence not encompassing the privacy protections of image subjects. Additionally, the legal focus on the collection of ``scans of face geometry'' without consideration of the identifiability of those scans points to possible privacy tensions in face blurring operations that may collect and utilize scans of facial geometry from images in pre-processing.
\\\\
Purpose of data use & The purpose for which data and FPT are used are central to determinations of the ethical valence of the technology. &BIPA regulates information and is agnostic as to how that information is used or how a particular technology built using the information is used.\\
\end{tabular}
\end{table}

\subsection{Control and Battery Consumption}\label{section:controlandbatteryconsumption}

It is generally presumed in the technical community that data analysis and machine learning model training that occurs ``locally'' or ``on-device'', such that the data never touches company servers, provide greater levels of privacy~\citep{googledevelopers, appleprivacy}. Accordingly, in lawsuits, companies have attempted arguments that, because their processing of biometric data is done locally, FPT regulations do not apply. However, FPT BIPA lawsuits have pointed to the importance of consumer control over the data processing that occurs on their devices, even if the data resulting from the processing may not touch company servers. Arguments about device battery consumption in legal complaints also introduce another dimension of consideration in technical on-device processing and privacy-preserving methods. Court opinions regarding data possession and the role of energy consumption in determining privacy harms can be contrary to commonly held values in the technical community, but can expand what it means for a technology or a technical method to be ``privacy-preserving''.

In \textit{Hazlitt v. Apple}, Apple claimed that, because face templates for face grouping in the photos application were only collected and processed locally on user devices, Apple did not possess the biometric data in a manner that could imply BIPA protections~\citep{apple061421}. This claim reflects narratives in the technical community that on-device processing broadly implies an increased level of privacy protection, perhaps, to the point where other privacy-preserving mechanisms such as consent and control are no longer needed. Accordingly, privacy-preserving and anonymization techniques have emphasized the obscuring of consumer data before that data is sent to servers~\citep{appledifferential,mcmahan2017federated}. For example, local processing is important in federated learning, a privacy-preserving machine learning technique, where, rather than having a model training data in a centralized server, data is decentralized across user devices. This technique has been used for privacy-preserving image classification~\citep{tensorflow}, and there is a proposal for the use of federated learning to train face recognition models~\citep{kim2021federated}. Snap also uses device-distributed machine learning and federated learning to power its lens feature~\citep{snapengineering2020}. 

However, the Disctrict Court for the Southern District of Illinois in  \textit{Hazlitt v. Apple} emphasized that, even if on-device data processing can provide greater privacy protections, it does not necessarily absolve that data of privacy protections that would otherwise be afforded to data processed on company servers. Citing an Illinois Supreme Court determination that possession ``occurs when a person has or takes control of the subject property or holds the property at her or his disposal'', the District Court in Apple’s case ruled that ``Plaintiffs have adequately alleged that Apple possessed their biometric data such that BIPA section 15(a) applies to it'' and did not dismiss the count~\citep{apple061421}. While the plaintiffs’ biometric data was not transmitted to company servers, they did not have control over whether biometric data was collected or whether face detection or template extraction was used for face grouping on their devices. They could not disable face grouping on their device, nor could they remove the ``People'' album. This court opinion that companies still possess data on consumer devices in a way that implicates responsibilities for privacy protections similarly reflects public outrage at the privacy intrusion over the lack of control that resulted from Apple’s since-delayed NeuralHash perceptual hashing model program, for which Apple similarly emphasized on-device processing and anonymization techniques but left users without control over how their data was handled~\citep{whittaker2021}. 

Another aspect of on-device processing that has been explored in privacy cases is battery consumption and device storage. Appealing to economic injuries to establish standing in courts, plaintiffs have attempted arguments about device battery consumption. In a complaint filed against TikTok for violation of multiple laws, including BIPA and the California False Advertising Law, the plaintiffs alleged that TikTok’s conduct led to harm to their mobile devices and incurred unnecessary electricity usage costs because of TikTok’s collection of personally identifiable and biometric image and video information~\citep{tiktok121820}. This complaint suggests the importance of cost considerations in data collection on-device, even when data collection and model training are ``privacy-preserving''. There are already hints of this consideration in private industry, with Snap acknowledging the importance of being ``mindful of device and network resources and the types of models that can be used given the constraints''~\citep{snapengineering2020}.

However, it has been suggested that arguments about battery consumption do not actually capture the essence of the actual privacy harms. Instead, arguments about battery consumption in privacy complaints arise out of perverse incentives for plaintiffs to cite harms that would be recognized as concrete~\citep{citron2022}. Certainly, the prioritization of device battery consumption as a privacy harm above all else may discourage on-device processing even where it can be privacy-preserving, but the point here is that data processing that occurs on user devices should not necessarily absolve the responsibility that comes with the possession or processing of the data. Even aside from energy consumption, Kearns and Roth~\citep{kearns2019} note that privacy-preserving techniques in industrial applications are generally utilized as an avenue to access new and more aggregations of user data, rather than retroactively applied to existing algorithms that run on user data to increase privacy protections. With anonymization techniques and on-device processing framed within industry as new ways to collect more data and touted to the public as privacy-preserving, these legal cases highlight that preserving privacy is a multidimensional problem, with on-device processing and anonymization techniques being approaches from just one axis. Court opinions from BIPA litigation underscore that technical approaches to privacy are not enough, surfacing other dimensions of privacy-related harms, such as the lack of user control over the features and data processing that are on their devices, as well as the lack of transparency over the energy costs of contributing their data to the company’s ML models.


\subsection{Consent in the Collection of Biometric Data}\label{section:consent}
Under BIPA, informed consent involves providing a written notice that a biometric identifier or biometric information is being collected and stored; providing the purpose and how long for which the information is being collected, stored, and used; and receiving a written release by the subject of the biometric information. Recent FPT litigation demonstrates BIPA’s governance of consent in the procurement of biometric data from public image sources, where subjects of biometric information do not have any relationship with the entity collecting it. This is of particular relevance in the computer vision community, within which preserving privacy in the procurement of data from public images is a critical ethical challenge. Table 1 in~\citep{prabhu2020} highlights how, in some of the most popularly used large-scale image datasets containing facial images, none of images were collected consensually. With the majority of surveyed scientists believing that images are permissible to collect if they are available online or based on licenses~\citep{noorden2020}, BIPA’s singular focus on informed consent for privacy-preserving FPT may not be similarly shared by the technical community.

Privacy and fairness in the development of FPT can be difficult to optimize for simultaneously~\citep{xiang2021}. While Buolamwini and Gebru~\citep{buolamwini2018} attributed the unfairness and disparities in the performance of FPT for darker-skinned, feminine-presenting individuals to a lack of sufficiently diverse training datasets featuring such individuals, as Xiang~\citep{xiang2021} discusses, with the need for millions of images in the development of FPT, from a large number of globally diverse data subjects is a non-trivial task that can easily go wrong and requires clearer regulatory guidance. Andrus et al.~\citep{andrus2021} also highlight the many privacy challenges to mitigate biases within AI systems, including challenges that come with obtaining consent for the procurement of demographic data to be able to check for diversity and bias in the first place. Xiang~\citep{xiang2021} also notes the differences between companies that can collect large image datasets from users who upload their images for free and provide consent in exchange for services versus companies that do not have users directly providing such data, highlighting possible competition issues. This suggests the need for investment in more open, consensual datasets that comply not only with privacy laws but also strong diversity requirements and fair compensation of data subjects. 

While usage of publicly available images is common in computer vision, including in efforts to develop less biased models, a string of BIPA lawsuits spurred by the release of IBM's Diversity in Faces dataset~\citep{merler2019} concluded that written consent is required even if the images are public and the companies do not have a relationship with the data subject. In \textit{Vance v. IBM}, IBM was sued for the non-consensual extraction of plaintiffs’ biometric data in the curation of IBM’s Diversity in Faces dataset~\citep{ibm091520}, which was released and curated by IBM in response to challenges in building fair FPT~\citep{merler2019}. Diversity in Faces consists of a selection of one million facial images from the YFCC100m dataset, which is a collection of over 99 million photographs from Flickr under a creative commons commercial or non-commercial license~\citep{thomee2016}. The court affirmed the plaintiffs’ allegations of IBM’s BIPA violations requiring written consent in the collection and dissemination of biometric information. Building off this lawsuit, in \textit{Vance v. Microsoft} and \textit{Vance v. Amazon}, Microsoft and Amazon were sued for using IBM’s Diversity in Faces dataset to train their FPT. Similarly, while the companies argued that the consent policy did not apply when there was not a direct relationship between the subject of the information and the company, the court overseeing these two cases had the opinion that consent under BIPA Section 15(b) is required ``when a private entity collects, captures, purchases, trades for, or gets biometric data in some other way''~\citep{amazon031521}, and that Microsoft and Amazon procured the data ``some other way'' through IBM’s dataset. 

But even with BIPA’s requirement of consent from subjects in public data sources, there remains uncertainty around how consent should and can be obtained from subjects in sources that historically have been treated as ``fair game'' for computer vision researchers. For example, for datasets that include the biometric information of subjects from Flickr images, the subjects of those images may not have any relationship with the publishers of the images. What is reasonable to ask companies to do to obtain consent from those subjects in these instances? Would those data sources simply be off-limits? What does that imply for past models developed using those images?

Other legal mechanisms that have been counted on to protect individuals' rights with regard to image datasets, such as copyright and licenses, have also not been the most effective. Copyright does not protect the privacy rights of subjects of the photograph, but the expression rights of the authors of the photograph. As Sobel~\citep{sobel2020} notes, what is useful to facial processing algorithms is usually not the portions of photographs that contain copyrightable artistic expressions, but clear, plain images of faces. So, the data that is useful to these algorithms is ``unrelated to the expression in a photograph that copyright protects''~\citep{sobel2020}. Peng et al.~\citep{peng2021} also suggest the ineffectiveness of licenses in prohibiting commercial use of datasets, noting the use of certain datasets in commercial settings against the curators' intentions.

\subsection{Identifiability}\label{section:identifiability}
BIPA litigation has surfaced two points in the realm of subject identifiability in FPT. First, because, for FPT, BIPA broadly protects ``scans of face geometry'',  the identifiability of a particular technology that utilizes scans of face geometry has not been considered, as long as scans of face geometry are implicated. Additionally, the identifiability of scans of face geometry themselves have not mattered, as long as they are scans--meaning, face templates and facial landmarks have so far been treated similarly. This might run counter to both BIPA itself, which considers these scans ``biometric \textit{identifiers}'' and emphasizes the potential for identity theft, as well as values in the technical community, which might emphasize the potential for identification. Second, while the technical community has made strides in techniques such as face replacement and obfuscation~\citep{yang2021, newton2005} to decrease the level of identifiability of subjects in images and videos, BIPA explicitly excludes protections for photographs, including facial photographs, and BIPA FPT litigation has not addressed subject identifiability in images. There is a contrasting lack of acknowledgement by courts of the degree of identifiability of a piece of technology or data and the harms resulting from the collection and distribution of identifying facial images. In this section, we elaborate on and discuss the implications of these two points for FPT development.

As long as scans of face geometry are collected and used, it has not been seriously considered in courts whether the scans or the particular technology resulting from the scans identifies or has the capability to identify individuals. Both Apple and Shutterfly claimed that their face grouping technologies are ``anonymous'' because the technology does not attempt to identify any faces, but to group similar faces~\citep{apple111220, shutterfly051921}. Shutterfly settled the lawsuit with \$6.75 million before the argument could be addressed~\citep{shutterfly051921}. However, the District Court did not agree with Apple that face grouping was anonymous because Apple still collected face templates, considered a scan of face geometry. But even when the underlying data is a scan of face geometry, the identifiability of the scan can vary. In TikTok’s settlement document, TikTok claimed that both the facial landmarks it collects and the demographic data generated from its landmarking technology are anonymous~\citep{tiktok121820}. The company emphasized that the landmarks are incapable of identifying individuals and are used to locate certain features of the face and to derive visual patterns that indicate demographic characteristics. It is generally the case that facial landmarks have fewer identifying capabilities than face templates, but TikTok’s argument was not explored further because TikTok settled the lawsuit~\citep{tiktok121820}.  This settlement might still speak to BIPA’s emphasis on general scans of face geometry and the lack of consideration the policy has for the actual identifying capabilities of the scans or the technology resulting from the scans. This is especially interesting given the focus of the policy’s legislative intent on identity theft and the uniqueness of biometric data to a particular individual.

Another feature of BIPA that can lead to identifiability tensions in the development of FPT is its explicit exclusion of photographs. In \textit{Vance v. IBM}, landmarks and other measurements were extracted from the plaintiff’s facial images and published alongside their facial images as part of the Diversity in Faces dataset. The plaintiff’s complaint in IBM’s case noted the privacy intrusion that came with the mass collection and distribution of their identifying facial images as well as biometric data~\citep{ibmcomplaint011420}. On the other hand, the court, in line with BIPA, focused on the biometric measurements that accompanied the images~\citep{ibm091520}. The explicit exclusion of photographs from BIPA’s definition of biometric identifiers, and, thus, BIPA protections, has placed legal focus on the extraction of biometric data from facial images, rather than on the mass collection and distribution of facial images themselves as part of image datasets. Indeed, in \textit{American Civil Liberties Union v. Clearview AI, Inc.}, the court noted that ``BIPA does not prohibit Clearview from collecting or republishing publicly-available photographs or expressing an opinion about who is pictured in them''~\citep{clearviewai082721}. 

While facial images in the context of FPT tend to be inextricably tied to biometric data extraction, images themselves are also identifying pieces of data. Even without face templates or facial landmarks attached to them, images are still identifying and can include sensitive information such as time and location data~\citep{andrews2021}. In image databases used for training computer vision technologies, after the images are selected and used as part of the image dataset, it is possible that the image is subsequently removed from the original source while it still lives on in the dataset. In fact, as Peng~\citep{peng2020} notes, images that are removed from training datasets can become even more widely distributed, as they are included in copies of datasets made by other researchers. Additionally, the harms that can come from the non-consensual collection and distribution of facial images versus the non-consensual collection and distribution of biometric information have also been addressed as distinct privacy harms in the literature~\citep{paullada2021} (though both can be imparted through FPT development and deployment). Perhaps in recognition of the distinct privacy intrusions that can come with the distribution of identifying facial images, other policies such as the CCPA~\citep{calegislature2018} and the EU General Data Protection Regulation (GDPR)~\citep{eu2016}  include facial images as part of their definitions of biometric data.

The exclusion of facial images from privacy protections in privacy policies for FPT can also present conflicting incentives in de-identification processes for the development of FPT. With BIPA being agnostic to both the identifiability of the technology resulting from the use of biometric identifiers and the biometric identifiers themselves, the bounds of its regulatory power for technical operations that could increase the level of privacy in image datasets become murky. 

One goal in the technical community is to allow researchers access to good quality image data while maintaining the privacy of faces in images. Towards this goal, ImageNet, one of the largest and most widely used image datasets for object detection (not for face detection or recognition tasks), was re-released with obfuscated faces~\citep{yang2021}. Yang et al.~\citep{yang2021} show that, even with face obfuscation, object detection algorithms trained on ImageNet still exhibit a high degree of accuracy for object-related tasks. The authors use Amazon Rekognition’s face detection API to detect faces and crowdsourcing to further refine their results. What is worth noting here, though, is that Amazon Rekognition’s DetectFaces operation detects faces by looking for ``key facial features such as eyes, nose, and mouth''~\citep{rekognition}. By default, the location of those features are returned when a call to the API is made, but they are not used in the annotations and the call is a non-persisting call, This means that data extracted or used during the function call is not stored~\citep{rekognition}. Here, bounding boxes from the API, which may use facial landmarks, are required in the blurring of faces in images. So, if facial landmarks are considered biometric identifiers, the collection or capturing of these facial features by private entities would require notification and consent from subjects under BIPA. Taken to the extreme, perhaps counterintuitively, facial obfuscation on images to increase the level of privacy for subjects in the images could potentially violate BIPA because of the facial landmarks extracted along the way. This points to a tension in ambiguity between privacy protections for subjects in facial images, which has become increasingly important within the technical community, and protecting the subjects’ scans of face geometry. But because facial images are not given any privacy protections under BIPA in the first place, it becomes unclear what the policy creates incentives for in this case.

The privacy intrusion that comes with the collection and distribution of identifying facial images has been addressed as part of FPT complaints and in the technical community, but has not been addressed in BIPA or BIPA litigation. It could be helpful, in laws that do include facial images as part of their definition of biometric data, to have clarity about privacy protections for images that have been blurred, or to provide incentives for companies to pursue privacy-preserving techniques for images. Additionally, the actual identifying capabilities of the facial data that have been grouped under ``biometric identifiers'' have yet to be seriously considered. Regulatory consideration of the identifying capabilities of different scans of face geometry and the different kinds of legal protections that should be afforded as a result can be helpful, perhaps lending different treatment to facial landmarks that outline someone’s eyes and lips just enough for virtual makeup application and the face templates used to unlock phones.

\subsection{Purpose of Data Use}\label{section:purposeofuse}
Unlike federal privacy laws that protect specific domains of information, such as the Health and Insurance Portability Act (HIPAA) and Family Educational Rights and Privacy Act (FERPA), BIPA protects types of biometric identifiers and information rather than the purpose or the context in which the information is used. While disclosure of the purpose of information use is required as part of BIPA’s consent policy, the purpose of use does not change the degree of protection that BIPA provides. However, as the survey in~\citep{kugler2019} suggests, people have very different attitudes based on how their biometric data is used. BIPA’s governance of information has allowed it to govern multiple steps in the building and deployment of a broad range of FPT, but the policy’s affordance of the same level of privacy protections to biometric identifiers and information no matter how the data is used contrasts significantly from surveyed attitudes, as well as attitudes in the technical community about how to reason about the contextual design of FPT.

Much of the discourse surrounding FPT harms is concerned with its use in high-stakes areas, such as in law enforcement, where the utilization of FPT has led to false arrests and bannings from commercial establishments. A combination of the efficiency with which potential face matches can be found with FPT and the inaccuracy of FPT on dark-skinned and/or feminine-presenting faces create breeding grounds for disaster. Additionally, aside from assessments of potential material harms, how people perceive the level of privacy intrusion introduced by FPT is also highly related to how it is used. For example, while people may generally be comfortable with the use of FPT to unlock their phones, they are generally uncomfortable with the use of FPT to surveil in public contexts~\citep{kugler2019}.

Similarly, there is an attitude among FPT researchers that the ethical risks associated with face filters differ from the risks associated with searchable face databases, with one researcher bringing up an implicit distinction that is made within the technical community between face filters and surveilling face identification technology—noting that ``no one’s threatening your Snapchat filter''~\citep{raji2020}. Except that, in some sense, people are. TikTok’s collection and use of facial landmarks from its users for face filters, stickers, and demographic classification without obtaining requisite consent landed the company in a 92-million-dollar BIPA settlement~\citep{tiktok121820}. In this situation, the major privacy intrusion may not be in the possession of the data itself or in the application of that data to impose face filters and stickers on users, but in the misuse of facial landmarks to perform unexpected and intrusive operations, such as demographic classification for advertisement targeting. But, under BIPA, policies do not differentiate between use cases of facial landmarks. In TikTok’s settlement document, the court suggests that it is the underlying data itself that gives rise to BIPA protections, not necessarily their applications~\citep{tiktok121820}. Given BIPA’s focus on information, just the underlying data in face filters and demographic classification is considered in determining the plausibility of BIPA protections, not how the FPT is used. The policy’s approach, then, differs quite substantially from the European Union’s proposed Artificial Intelligence Act, which has separate policies based on whether the use of an AI system is ``high-risk'' or ``low-risk''. Under the EU’s proposed AI Act, the use of AI systems that detect demographic information using biometric data, or ``biometric categorisation systems'', must be disclosed~\citep{eu2021}. It might govern TikTok’s use of biometric data to categorize its users differently from its use of biometric data to apply face filters, more aptly capturing the privacy intrusion that is described in~\citep{garvie2016}. Just as Selbst et al.~\citep{selbst2021} encourages the widening of abstraction boundaries in the development of facial recognition technology to encompass the contexts under which they are deployed, we also encourage a policy approach that considers the context under which biometric data or FPT are used.

\section{Technology and Policy Implications}\label{section:techpolicy}

With private and public actors building FPT in an increasing number of settings, these privacy tensions imply several implications for FPT policies and FPT development moving forward. Firstly, it is important for FPT policies to recognize protections for subjects of both facial images and biometric data, and possibly a consideration of the intersection of these protections, such as cases where the obfuscation of faces on a large scale require the extraction of biometric data. Here, it may be helpful to acknowledge privacy-preserving techniques for both forms of data, such as the encryption of biometric data and face obfuscation in images. For example, ways in which other policies have created incentives for the employment of privacy-preserving techniques have been through carve-outs for when data is anonymized under GDPR~\citep{eu2016} or encrypted under CCPA~\citep{calegislature2018}. The effects of Illinois BIPA’s governance of FPT thus far have generally just been companies just not offering certain biometric features in Illinois~\citep{langone2018, sony2018}, rather than an effort to comply with the policy in the development of FPT. There is not a clear path that the policy offers companies who do want to build and deploy technologies in Illinois that incorporate biometric data. On the other hand, under the GDPR, many companies have invested in anonymization techniques, and there have been several large-scale implementations of anonymization libraries~\citep{guevara2020,bird2020}.  Meanwhile, it is important for FPT developers to keep privacy values beyond purely technical notions of anonymization in mind. These considerations include user control over data, even if all data processing happens on the device, and, perhaps, even the consideration of battery consumption involved in the processing of data for a purpose that the user did not consent to. 

Second, it is important to critically examine the role of identifiability in FPT policies. High levels of identifiability and identity theft should not be the only markers of a cognizable privacy intrusion in the context of FPT. For instance, facial landmarks, which typically do not provide as much identifiability information as face templates, can also be used in emotion recognition and to infer sensitive characteristics such as race and gender. But, where identifiability harms are present, it can be important to address the feasibility of identification from different forms of FPT and data used for FPT, e.g., the difference in identifiability levels in face templates versus facial landmarks. Next, there should be a consideration of the settings and contexts under which FPT are deployed. Public attitudes towards FPT suggest that there is less of a privacy intrusion with FPT applied onto oneself, such as the use of facial information to log into a bank account or the application of face filters, versus the collection of facial information through surveillance in a commercial establishment or on a public street~\citep{kugler2019}. This difference in the level of privacy intrusion when FPT are applied to oneself versus when FPT are used by someone else may also be captured in the EU AI Act’s emphasis on \textit{remote} biometric identification~\citep{eu2021}.

Finally, FPT policies should  provide  meaningful guardrails in the procurement of biometric training data. Determinations of particularized harms in the context of BIPA's profit clause bring to light a lack of recognition of the value and contribution of biometric training data for FPT, as well as the growing market for biometric training data. For FPT policy to provide these guardrails, there must be a recognition of the fundamentality of training data to FPT functionality and the privacy intrusions that can come, not only with the dissemination of biometric data in the final product, but also the procurement of that data in the development of FPT.

\section{Conclusion}\label{section:conclusion}
An understanding of privacy values for FPT in technical and legal communities is crucial to the development of well-informed technology policies and the development and deployment of FPT that are aligned with human norms more broadly. In this paper, we establish and explore Illinois BIPA’s role as a major governing mechanism for FPT in the United States, point out four disconnects in considerations between technical and legal communities that have arisen out of BIPA litigation and suggest implications for future policy design and FPT development. We found that (i) while the technical community has placed significant privacy value in on-device processing, the legal community has emphasized control over data even when it is completely processed on-device and in fact treats the battery power used in on-device processing as an economic harm; (ii) while the technical community has made strides in de-identification techniques for FPT, such as for face obfuscation, there is not an accompanying legal acknowledgement of these advances in BIPA litigation; (iii) while the legal community has emphasized informed consent as an important value in privacy-preserving FPT through BIPA, the technical community's implementation of informed consent has been very mixed, especially when pursuing other values like fairness; and (iv) while the technical community has placed an emphasis on the purpose of data use in determining levels of privacy protections, BIPA has treated the collection of data for all use cases similarly. Addressing these disconnects will be critical in the development of policies and technologies that are, taken together, oriented towards the mitigation of harms.

\begin{acks}
We thank Jerone Andrews and Dora Zhao for their terrific insight and their support throughout this research project. We are also grateful to Berhard Egger and Leilani Gilpin for clarifications about the identifiability of different forms of facial data, as well as to Kaiyu Yang for answering questions about ImageNet obfuscation. Finally, we thank Christian Cmehil-Warn for helpful discussions, Rebecca Spiewak for her valuable edits, and the anonymous reviewers for their constructive feedback.
\end{acks}

\bibliographystyle{ACM-Reference-Format}

\end{document}